\documentclass{emulateapj}
\usepackage{natbib}
\usepackage{apjfonts}

\shorttitle{THE $M_{\rm BH} - L_{\rm bulge}$ RELATIONSHIP FOR AGNs}
\shortauthors{BENTZ ET AL.}

\received{}
\accepted{}

\begin{document}

\title{The Black Hole Mass -- Bulge Luminosity Relationship for
Active Galactic Nuclei from Reverberation Mapping and Hubble Space
Telescope Imaging}

\author{ Misty~C.~Bentz\altaffilmark{1,2}, 
         Bradley~M.~Peterson\altaffilmark{2,3},
	 Richard~W.~Pogge\altaffilmark{2,3}, 
	 Marianne~Vestergaard\altaffilmark{4} 
}

\altaffiltext{1}{Present Address:
                 Department of Physics and Astronomy,
		 4129 Frederick Reines Hall,
		 University of California, Irvine,
		 Irvine, CA 92697;
		 mbentz@uci.edu}

\altaffiltext{2}{Department of Astronomy, 
		The Ohio State University, 
		140 West 18th Avenue, 
		Columbus, OH 43210; 
		peterson, pogge@astronomy.ohio-state.edu}

\altaffiltext{3}{Center for Cosmology and AstroParticle Physics,
                 The Ohio State University,
                 191 West Woodruff Avenue,
                 Columbus, OH 43210}

\altaffiltext{4}{Department of Physics and Astronomy, 
                 Robinson Hall, Tufts University, 
		 Medford, MA 02155;
		 M.Vestergaard@tufts.edu }

\begin{abstract}

We investigate the relationship between black hole mass and bulge
luminosity for AGNs with reverberation-based black hole mass
measurements and bulge luminosities from two-dimensional
decompositions of {\em Hubble Space Telescope} host galaxy images.  We
find that the slope of the relationship for AGNs is $0.76 - 0.85$ with
an uncertainty of $\sim 0.1$, somewhat shallower than the $M_{\rm BH}
\propto L^{1.0 \pm 0.1}$ relationship that has been fit to nearby
quiescent galaxies with dynamical black hole mass measurements.  This
is somewhat perplexing, as the AGN black hole masses include an
overall scaling factor that brings the AGN $M_{\rm BH} -
\sigma_{\star}$ relationship into agreement with that of quiescent
galaxies.  We discuss biases that may be inherent to the AGN and
quiescent galaxy samples and could cause the apparent inconsistency in
the forms of their $M_{\rm BH} - L_{\rm bulge}$ relationships.

\end{abstract}

\keywords{galaxies: active --- galaxies: nuclei --- galaxies: photometry 
--- galaxies: Seyfert}

\section{Introduction}

Most galactic bulges are now believed to harbor a massive black hole.
For AGNs, the evidence of the black hole is obvious from the activity.
However, even in quiescent galaxies, the effect of the black hole can
be detected from stellar and gas kinematics near the nucleus.  What
remains to be determined is the formation mechanism for these massive
black holes and their role in shaping and responding to the evolution
of their host galaxies.

In an early review on the subject of quiescent galaxy black hole
masses, \citet{kormendy95} pointed out that the estimated central
black hole masses of eight galaxies seemed to show a correlation with
the host galaxy bulge luminosities (or, equivalently, bulge stellar
masses).  \citet{magorrian98} later investigated a sample of 32 nearby
galaxies and their black hole masses, and confirmed that the estimated
black hole mass in each galaxy was indeed proportional to the
luminosity (or mass) of the host galaxy bulge, albeit with a scatter
about the fit of approximately $\pm 0.5$\,dex.  Later studies claimed,
in some cases, that there may be a difference in the black hole--bulge
relationship for quiescent and active galaxies, or even for Seyfert
galaxies and quasars (e.g., \citealt{wandel99a}).  Subsequent
investigations seem to have decreased these discrepancies through more
sophisticated techniques of measuring the bulge luminosity such as
2-dimensional image decompositions (e.g., \citealt{mclure01,wandel02}
[hereafter W02]) or dynamical modeling of the host galaxy (e.g.,
\citealt{haring04}).

We have recently completed a {\em Hubble Space Telescope} campaign to
image the host galaxies of AGNs with black hole masses from
reverberation-mapping.  The images were acquired for the purpose of
decomposing the surface brightness profiles of the host galaxies and
creating ``nucleus-free'' images from to measure the host-galaxy
starlight contributions to ground-based spectroscopic luminosity
measurements of the AGNs (results described by
\citealt{bentz06a,bentz08b}).  Combining the bulge luminosities
estimated from the surface brightness decompositions of these
high-resolution images with the recently updated and homogeneously
analyzed database of reverberation masses for these objects
(\citealt{peterson04,grier08}), we re-examine the relationship between
black hole mass and host galaxy bulge luminosity for nearby AGNs.
Throughout this work, we will assume a standard flat $\Lambda$CDM
cosmology with $\Omega_{\rm B} = 0.04$, $\Omega_{\rm DM} = 0.26$,
$\Omega_{\Lambda} = 0.70$, and $H_0 = 70$\,km\,s$^{-1}$\,Mpc$^{-1}$.

\begin{deluxetable*}{lccccccc}
\tablecolumns{8}
\tablewidth{500pt}
\tablecaption{Black Hole Masses and Bulge Luminosities}
\tablehead{
\colhead{Object} &
\colhead{$D_L$} &
\colhead{$E(B-V)$} &
\colhead{$m_{HST}$} &
\colhead{$m_V$} &
\colhead{$M_V$} &
\colhead{$\log L$} &
\colhead{$\log M_{\rm BH}$}\\
\colhead{} &
\colhead{(Mpc)} &
\colhead{(mag)} &
\colhead{(vegamag)} &
\colhead{(vegamag)} &
\colhead{(mag)} &
\colhead{($L_{\odot}$)} &
\colhead{($M_{\odot}$)}}

\startdata
Mrk\,335      &	113  &  0.035  & 16.16  & 16.26  &  $-19.00$  &  9.53	      & $7.15 \pm 0.11$  \\
PG\,0026+129  &	672  &  0.071  & 16.21  & 16.23  &  $-22.91$  &  11.09       & $8.59 \pm 0.11$  \\
PG\,0052+251  &	740  &  0.047  & 17.81  & 17.92  &  $-21.42$  &  10.50       & $8.57 \pm 0.09$  \\
Fairall\,9    &	209  &  0.027  & 15.13  & 15.20  &  $-21.40$  &  10.49       & $8.41 \pm 0.10$  \\
Mrk\,590      &	115  &  0.037  & 15.59  & 15.68  &  $-19.63$  &  9.78, 9.99  & $7.68 \pm 0.07$  \\
3C\,120	      &	145  &  0.297  & 16.64  & 16.72  &  $-19.08$  &  9.57	      & $7.74 \pm 0.21$  \\
Ark\,120      &	142  &  0.128  & 14.50  & 14.59  &  $-21.17$  &  10.40       & $8.18 \pm 0.06$  \\
Mrk\,79	      &	96.7 &  0.071  & 15.84  & 15.95  &  $-18.98$  &  9.52, 10.19 & $7.72 \pm 0.12$  \\
PG\,0804+761  &	461  &  0.035  & 16.70  & 16.74  &  $-21.58$  &  10.56       & $8.84 \pm 0.05$  \\
PG\,0844+349  &	287  &  0.037  & 16.75  & 16.81  &  $-20.48$  &  10.13       & $7.97 \pm 0.18$  \\
Mrk\,110      &	155  &  0.013  & 18.08  & 18.16  &  $-17.79$  &  9.05	      & $7.40 \pm 0.11$  \\
PG\,0953+414  &	1170 &  0.013  & 17.64  & 17.82  &  $-22.53$  &  10.94       & $8.44 \pm 0.09$  \\
PG\,1211+143  &	368  &  0.035  & 17.16  & 17.21  &  $-20.62$  &  10.18       & $8.16 \pm 0.13$  \\
PG\,1226+023  &	758  &  0.021  & 15.49  & 15.61  &  $-23.79$  &  11.45       & $8.95 \pm 0.09$  \\
PG\,1229+204  &	283  &  0.027  & 17.17  & 17.23  &  $-20.03$  &  9.94, 9.98  & $7.86 \pm 0.21$  \\
PG\,1307+085  &	740  &  0.034  & 16.66  & 16.70  &  $-22.65$  &  10.99       & $8.64 \pm 0.12$  \\
Mrk\,279      &	133  &  0.016  & 16.13  & 16.22  &  $-19.40$  &  9.69, 10.00 & $7.54 \pm 0.11$  \\
PG\,1411+442  &	410  &  0.008  & 16.70  & 16.74  &  $-21.32$  &  10.46       & $8.65 \pm 0.14$  \\
PG\,1426+015  &	395  &  0.032  & 15.36  & 15.34  &  $-22.64$  &  10.99       & $9.11 \pm 0.13$  \\
Mrk\,817      &	138  &  0.007  & 17.62  & 17.71  &  $-17.99$  &  9.13	      & $7.69 \pm 0.07$  \\
PG\,1613+658  &	606  &  0.027  & 16.14  & 16.20  &  $-22.71$  &  11.02       & $8.45 \pm 0.20$  \\
PG\,1617+175  &	522  &  0.042  & 16.36  & 16.34  &  $-22.25$  &  10.83       & $8.77 \pm 0.10$  \\
PG\,1700+518  &	1510 &  0.035  & 17.54  & 17.67  &  $-23.22$  &  11.22       & $8.89 \pm 0.10$  \\
3C\,390.3     &	251  &  0.071  & 16.74  & 16.80  &  $-20.19$  &  10.01       & $8.46 \pm 0.10$  \\
Mrk\,509      &	151  &  0.057  & 13.96  & 13.96  &  $-21.94$  &  10.71       & $8.16 \pm 0.04$  \\
PG\,2130+099  &	283  &  0.044  & 18.73  & 18.79  &  $-18.47$  &  9.32	      & $7.58 \pm 0.17$  \\
											      
\enddata

\tablecomments{For those objects with two bulge luminosities listed,
               the first is the luminosity of the largest-scale
               non-disk component, and the second is the luminosity of
               all non-disk components, including bars or inner
               bulges.  Black hole masses are from \citet{peterson04},
               except for PG\,2130+099 \citep{grier08}.}

\end{deluxetable*}

\section{Bulge Luminosities and Black Hole Masses}

We investigate here the sample of AGNs with black hole masses measured
using the variability technique known as reverberation mapping.  The
details of the host galaxy imaging and surface brightness
decompositions are described by \citet{bentz06a,bentz08b}, but we
include a short summary of the relevant details here.  The majority of
the objects in the reverberation sample were imaged with the Advanced
Camera for Surveys (ACS) High Resolution Channel (HRC) through the
F550M filter.  Unfortunately, ACS ceased functioning before the
observations were completed, so five objects (PG\,0026, PG\,1307,
PG\,1426, PG\,1617, and Mrk\,509) were imaged with the Wide Field
Planetary Camera 2 (WFPC2) through the F547M filter.  Exposure times
for all observations were graduated so that the saturated pixels in
the nucleus region of the long exposures could be corrected using an
unsaturated image.  After correcting saturated pixels and cleaning
cosmic rays, individual images were co-added and corrected for
distortion if necessary.  The surface brightness profiles of the AGN
host galaxies were modeled with the two-dimensional image
decomposition program Galfit \citep{peng02}.  To avoid overcorrecting
the AGN luminosities for host galaxy starlight, we made conservative
host galaxy model fits which may somewhat underestimate the true
brightness of the various galaxy components.

For most of the reverberation-mapped objects, the {\em HST} images are
sufficient for a full decomposition.  However, for the nearby and
spatially-extended NGC objects in the sample, the {\em HST} images do
not provide the necessary wide field of view (FOV) to measure the sky
background, which affects the accuracy to which we can constrain the
bulge and disk parameters.  Our current surface brightness models for
the {\em HST} images of the NGC galaxies include components that
cannot be resolved in the ground-based images due to their small
effective radii (<3.5\arcsec\ in all cases).  The typical seeing in
the ground-based images of $\sim 2$\arcsec\ blurs the central PSF
together with these compact components and with the bulge itself,
which has an effective radius of only $\sim 10$\arcsec\ for these
objects.  The difficulties in disentangling the nuclear structure in
the ground-based images and the fact that the distortion-corrected HRC
FOV is, at best, only twice the effective radius of the bulge and
furthermore doesn't allow measurement of the sky, creates enough
uncertainty in the bulge parameters for the NGC objects with
reverberation masses that we exclude them from further analysis at
this time.  Additional images with an intermediate FOV are necessary
to fully constrain the decompositions of these galaxies.  We also
exclude IC\,4329A because of its uncertain black hole mass and because
it is a dusty edge-on galaxy with an unreliable surface brightness
decomposition.  Table~1 lists the 26 objects examined by this paper.

The parameters of the surface brightness decompositions are included
in \citet{bentz08b}.  We take the component with the largest effective
radius (other than any exponential disk component) as the ``bulge''
for each of these galaxies, except for Mrk\,79 where there is clearly
an extended bar-like component.  For ellipticals, the bulge magnitude
is the total galaxy magnitude.  We followed the prescriptions of
\citet{sirianni05} for converting from the space telescope
AB$_{\lambda}$ magnitude system to Galactic extinction-corrected
magnitudes in the Vega system, which we list in Table~1 as $m_{HST}$.
The $V - HST$ color for each object was calculated with {\em synphot}
in IRAF using the bulge template of \citet{kinney96}.  We include in
Table~1 the extinction-corrected integrated bulge magnitude in the
Vega system for each AGN host galaxy, the corresponding apparent and
absolute Johnson $V$ magnitudes, and the $V$-band luminosity from the
standard relation $\log L_{\rm V}/L_{\odot} = 0.4(-M_{\rm V} + 4.83)$.
We assume an uncertainty of $\pm 0.2$\,dex for the bulge luminosities.
Four of the galaxies had an additional surface brightness component
besides a bulge or exponential disk.  Imaging alone is not sufficient
to ascertain whether these components contribute to the total bulge
brightness or are dynamically distinct, so we also list in Table~1 the
bulge luminosity for those objects including the contribution from the
additional component.

Figure~1 compares the (single component) bulge luminosities and black
hole masses for the sample of objects that are common to both this
work (filled circles) and the \citet{wandel02} study (open circles).
There is a clear trend between black hole mass and bulge luminosity
for the objects in this study, whereas the bulge luminosities from
\citet{wandel02} have a range of $\sim 2$\,dex in black hole mass over
a limited $\sim 1.5$\,dex range in luminosity.  It is perhaps
unsurprising that a correlation is more apparent now, as the study by
\citet{wandel02} is a compilation of inhomogeneous data from the
literature.  The bulge luminosities were primarily from
\citet{mclure01} and include measurements from ground-based
photographic plates and CCD imaging, as well as saturated and
unsaturated {\it HST} imaging, all in various passbands.  Many of the
studies included in the compilation did not use the same cosmologies
and distances for the objects or the same analysis techniques in
determining the black hole mass.  All of the reverberation-based black
hole masses have since been homogeneously analyzed and updated by
\citet{peterson04}, and PG\,2130+099 which has an updated mass from
the analysis of a new reverberation data set by \citet{grier08}.
Figure~2 shows the black hole masses versus the host galaxy $V$-band
luminosities for the full sample of 26 objects included in this study.

\begin{figure}
\epsscale{1.2}
\plotone{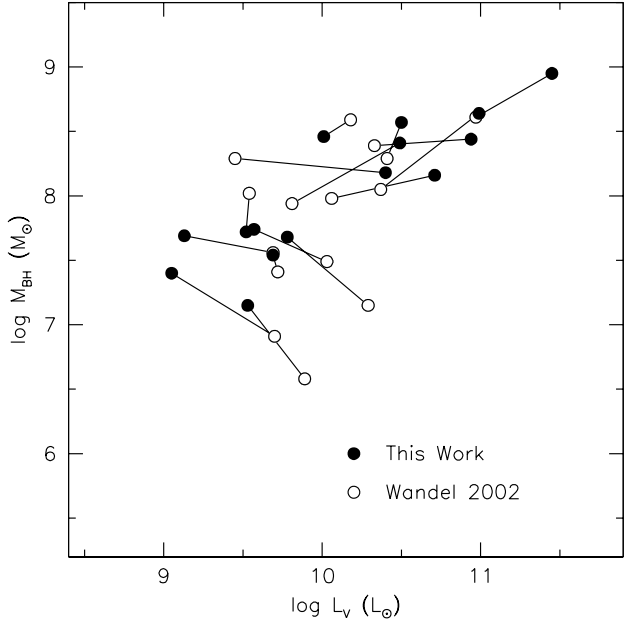}
\caption{Comparison of black hole mass and $V$-band bulge luminosity
         values from \citet{wandel02} ({\it open points}) and this
         work ({\it filled points}) for objects that are common to
         both.  The new values cover a range of 2.5\,dex in luminosity
         and show a clear trend, while the values from
         \citet{wandel02} have a range of $\sim 2$\,dex in mass within
         a limited $\sim 1.5$\,dex range in luminosity.}
\end{figure}

\section{The Black Hole Mass -- Bulge Luminosity Relationship}

We employed two independent fitting routines in our examination of the
$M_{\rm BH} - L_{\rm bulge}$ relationship: FITEXY \citep{press92},
which estimates the parameters of a straight-line fit through the data
including errors in both coordinates; and BCES \citep{akritas96},
which accounts for the effects of errors on both coordinates in the
fit using bivariate correlated errors and a component of intrinsic
scatter.  FITEXY numerically solves for the minimum orthogonal
$\chi^2$ using an interative root-finding algorithm and is a
``symmetric'' algorithm in that it does not assume a dependent and
independent variable.  Following \citet{tremaine02}, we include an
estimate of the fractional scatter, in this case the fraction of the
$M_{\rm BH}$ measurement value (not the error value) that is added in
quadrature to the error value to obtain a reduced $\chi^2$ of 1.0.
While BCES also accounts for intrinsic scatter, it does not provide a
measure of it.  We adopt the bootstrap of the BCES bisector value with
$N=1000$ iterations.  Fits of the form

\begin{equation}
\log \frac{M_{\rm BH}}{10^8 M_{\odot}} = K + \alpha \log \frac{L_{\rm bulge}}{10^{10} L_{\odot}}
\end{equation}

\noindent were performed utilizing both the single-component bulge
luminosities and the multiple-component bulge luminosities and are
presented in Table~2.  The powerlaw slope ranges from $0.76 \pm 0.08$
to $0.85 \pm 0.11$ depending on the definition of the bulge luminosity
and the specific fitting routine utilized.  We take the BCES fit
nearest the middle of this range, with a slope of $0.80 \pm 0.09$, as
the ``best'' fit.

\begin{figure}
\epsscale{1.2}
\plotone{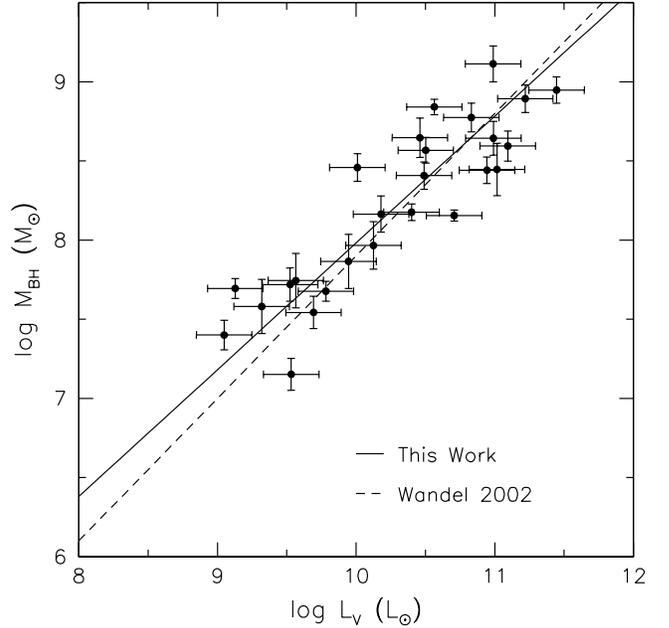}
\caption{The $M_{\rm BH} - L_{\rm bulge}$ relationship for AGNs with
         reverberation-based masses and bulge luminosities from
         two-dimensional decompositions of {\em HST} host-galaxy
         images.  The solid line is the BCES fit with a slope of
         $\alpha = 0.80 \pm 0.09$.  The dashed line is the fit from
         \citet{wandel02} to his sample of broad-line AGNs and has a
         slope of $\alpha = 0.90 \pm 0.11$.}
\end{figure}

For comparison, we fit the quiescent galaxy $M_{\rm BH} - L_{\rm
bulge}$ relationship using the sample of nearby galaxies with
dynamical black hole mass measurements (\citealt{ferrarese05}; FF05).
We restricted the sample to ellipticals, both to circumvent the need
for bulge--disk decompositions and because ellipticals are reported to
show less scatter about the $M_{\rm BH} - L_{\rm bulge}$ relationship
(cf. \citealt{mclure01}).  Bulge magnitudes were converted to $V$-band
using a typical elliptical galaxy color of $B-V=0.9$.  The fitting
results are presented in Table~2.  We also fit the quiescent galaxy
relationship excluding Cygnus\,A and NGC\,5845, both of which are
known to deviate significantly (FF05).

Figure~3 shows the $M_{\rm BH} - L_{\rm bulge}$ relationship for the
FF05 ellipticals compared to the ``best'' fit for the AGNs in this
study.  The slope of the quiescent galaxy relationship is steeper,
although the severity of the discrepancy depends on the specifics of
the fitting routine used, as the best-fit slope from BCES is $\alpha =
1.43 \pm 0.21$ versus $\alpha = 1.11 \pm 0.21$ from FITEXY.  The fits
presented here for the quiescent elliptical galaxies are slightly
steeper than that reported by FF05 ($\alpha = 1.05 \pm 0.21$) and are
in reasonable agreement with the slope of $1.40 \pm 0.17$ for the
quiescent galaxy fit reported by \citet{lauer07a}.  Both of these
studies included lenticular and spiral galaxies with dynamical masses
in their fitting samples.  \citet{lauer07a} compiled decompositions
from the literature, while FF05 assumed specific ratios of $L_{\rm
bulge}/L_{\rm total}$ based on morphological type.\footnote{The
relative contribution of the bulge is known to vary substantially
within a single morphological type.  Figure~6 of \citet{kent85} shows,
for example, that $B/T$ ranges from $0.3 - 1.0$ for S0 galaxies).}

\section{Discussion}

The compendium of AGN black hole masses and bulge luminosities
presented here offers several advantages over previous measurements
compiled from the literature (i.e.,
\citealt{wandel99a,wandel02,mclure01}).  The masses result from a
homogeneous analysis of reverberation-mapping data \citep{peterson04},
in contrast to many recent studies of AGN black hole -- bulge
relationships where black hole masses are inferred from single-epoch
spectral measurements (e.g., \citealt{greene08,kim08}).  The bulge
luminosities in this work are estimated from two-dimensional surface
brightness decompositions of unsaturated high-resolution space-based
images taken with the same instrument and the same filter.  The
exceptions are the five objects that were imaged through the F547M
filter using WFPC2, but the bandpass is very similar to the ACS F550M
filter employed for the other objects ($\lambda_{\rm c} {\rm
(F550M)}=5580$\,\AA\ versus $\lambda_{\rm c}{\rm (F547M)}=5483$\,\AA,
and $\Delta \lambda{\rm (F550M)} = 547$\,\AA\ versus $\Delta
\lambda{\rm (F547M)}=483$\,\AA).

\begin{deluxetable}{lccc}
\tablecolumns{4}
\tablewidth{220pt}
\tablecaption{Fits to the $M_{\rm BH} - L_{\rm bulge}$ Relationship}
\tablehead{
\colhead{Sample} &
\colhead{$K$} &
\colhead{$\alpha$} &
\colhead{Scatter\tablenotemark{a}}}

\startdata

\multicolumn{4}{c}{BCES} \\ \hline \\
AGNs                            & $-0.02 \pm 0.06$  & $0.80 \pm 0.09$  & \\
AGNs (+ extra components )      & $-0.07 \pm 0.08$  & $0.85 \pm 0.11$  & \\
FF05 Ellipticals                & $0.42 \pm 0.12$   & $1.43 \pm 0.21$  & \\
FF05 Ellipticals ($-$ outliers) & $0.30 \pm 0.10$   & $1.42 \pm 0.24$  & \\

\\
\hline \\
\multicolumn{4}{c}{FITEXY} \\ \hline \\

AGNs                            & $-0.05 \pm 0.06$  & $0.76 \pm 0.08$  & 0.38 \\
AGNs (+ extra components)       & $-0.13 \pm 0.06$  & $0.80 \pm 0.09$  & 0.44 \\
FF05 Ellipticals                & $0.15 \pm 0.11$   & $1.11 \pm 0.21$  & 0.73 \\
FF05 Ellipticals ($-$ outliers) & $0.14 \pm 0.11$   & $1.18 \pm 0.19$  & 0.64 \\

\enddata

\tablenotetext{a}{The fractional scatter, quantified as the fraction
                  of the measurement value of $M_{\rm BH}$ that must
                  be added in quadrature to the error value in order
                  to obtain a reduced $\chi^2$ of $1.0$.}

\end{deluxetable}

Unfortunately, there is not a similarly consistent sample of
high-quality images from which the bulge luminosities can be estimated
for the quiescent galaxies with dynamical black hole masses.  While
high-quality observations do exist for these nearby and well-studied
galaxies, they have not been obtained in a uniform fashion.  The
recent work by \citet{graham07} attempts to compensate for the
different methods and analysis techniques employed in several
publications (\citealt{mclure02b} with an updated cosmology presented
in \citealt{mclure04,marconi03,erwin04}), all of which arrive at
different values for the slope of the quiescent galaxy $M_{\rm BH} -
L_{\rm bulge}$ relationship and/or different black hole masses
predicted for a specific bulge luminosity. \citeauthor{graham07}
carefully updated and revised the samples of objects included in these
studies and found that the differences of the best-fit parameters
found by each study are mitigated, the scatter in the measurements is
significantly decreased, and that $M_{\rm BH} \propto L^{1.0}$.
Interestingly, he finds a somewhat shallower slope, $\alpha \approx
0.75 - 0.88$, when bulge luminosities are estimated by two-dimensional
decompositions of $B$-band images and corrected for
inclination-dependent dust extinction in the host galaxy disks.  Only
13 objects are included in that particular analysis (an updated form
of the study presented by \citealt{erwin04}) but it is intriguing
nonetheless in its close agreement with the fit that we find for the
AGNs.

While we expect that the AGN bulge luminosities in this may be
slightly underestimated based on the conservative galaxy fits we
employed, there is also a small $k$-correction introduced, as the
average redshift of the 26 AGNs is $z \approx 0.1$.  The portion of
the SED observed through the medium-band $V$ filters employed in the
{\em HST} observations is fainter ($\sim 0.1$\,dex in luminosity) than
if the galaxies were at $z=0$, assuming their stellar populations
resemble that of the bulge template of \citet{kinney96}.  Accounting
for any such biases in the galaxy fitting or colors would intensify
the apparent differences between the slopes measured for the AGNs and
quiescent galaxies.  The black hole masses for the AGNs have already
been scaled so that their $M_{\rm BH} - \sigma_{\star}$ relationship
is brought into agreement with the quiescent $M_{\rm BH} -
\sigma_{\star}$ relationship fit to the same sample of quiescent
galaxies included in the FF05 study.  The differences may be mitigated
if \citet{marconi08} are correct in their suggestion that neglecting
radiation pressure leads to systematic under-estimation of black hole
masses from reverberation mapping data (see, however,
\citealt{netzer08}).

\begin{figure}
\epsscale{1.2}
\plotone{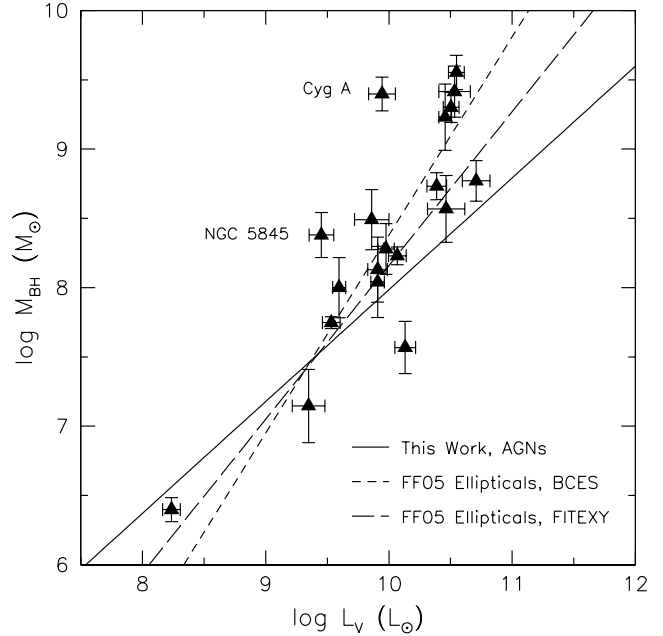}
\caption{The $M_{\rm BH} - L_{\rm bulge}$ relationship for quiescent
         galaxies as determined for the elliptical galaxies from
         \citeauthor{ferrarese05} (2005; filled triangles) compared to
         the $M_{\rm BH} - L_{\rm bulge}$ relationship for AGNs from
         this work.  The slope of the relationship appears shallower
         for AGNs than for quiescent galaxies.  The solid line is the
         same fit displayed in Figure~2 for the sample of AGNs in this
         work, and has a slope of $\alpha = 0.80 \pm 0.09$.  The
         short-dashed line is the BCES fit to the elliptical galaxies
         and has a slope of $1.43 \pm 0.21$, while the long-dashed
         line is the FITEXY fit to the elliptical galaxies and has a
         slope of $1.11 \pm 0.21$. }
\end{figure}

A separate concern has been raised by \citet{yu02},
\citet{bernardi07}, \citet{lauer07a}, and \citet{tundo07}, who present
an apparent disagreement between the quiescent galaxy $M_{\rm BH} -
\sigma_{\star}$ and $M_{\rm BH} - L_{\rm bulge}$ relationships and
suggest that the quiescent galaxy sample is biased towards galaxies
with overly large velocity dispersions for their luminosities (see,
however, \citealt{graham08} who examines the role of bars in this
issue).  There is no reason to suspect the AGN sample of having the
same bias, as the mass measurements are made using flux variability
techniques and not dynamical techniques.  Such a bias may help explain
why some of the quiescent galaxies at the high luminosity end have
black hole masses that are more than an order-of-magnitude larger than
the active galaxies, although it may not completely resolve this
disparity.

Finally, there may be no reason to expect that the $M_{\rm BH} -
L_{\rm bulge}$ relationship is the same for the AGNs and quiescent
galaxies in these samples, as there is only a modest number of objects
in each sample and selection effects likely play an important role on
both sides \citep{lauer07b}.  Clearly, there remain several areas that
are in need of investigation, any of which may shed light on the
apparently inconsistent fits to the $M_{\rm BH} - L_{\rm bulge}$
relationship for AGNs and for quiescent galaxies.  As the $M_{\rm BH}
- L_{\rm bulge}$ relationship is an important and widely used means of
estimating black hole masses throughout cosmic history (e.g.,
\citealt{marconi04,shankar04}), an accurate characterization of this
relationship is necessary for understanding black hole growth and
evolution as well as the interplay between black holes and their host
galaxies.

\section{Summary}

We have presented an updated version of the AGN $M_{\rm BH} - L_{\rm
bulge}$ relationship using the database of homogeneously analyzed
reverberation masses from \citet{peterson04} and \citeauthor{grier08}
(2008; PG\,2130+099) and the two-dimensional surface brightness
decompositions of the AGN host galaxies described by \citet{bentz08b}.
We find a strong correlation about the relationship for the 26 AGNs
included here, with a best-fit powerlaw slope of $0.80 \pm 0.09$.
This is somewhat shallower than the best-fit slope for quiescent
galaxies ($\alpha \approx 1.0$), even though the AGN black hole masses
have been scaled to bring the AGN and quiescent galaxy $M_{\rm BH} -
\sigma_{\star}$ relationships into agreement.  There appear to be many
systematics in both the AGN and quiescent galaxy samples that must be
investigated in order to more completely understand this important
relationship.

Our future plans include investigating the biases in the AGN sample
and extending the range of the relationship for AGNs.  We have an {\em
HST} Cycle 17 program to image the NGC objects that were excluded from
this particular work with the Wide Field Camera 3 through the F547M
filter.  These observations will provide us with the intermediate FOV
images necessary for accurate decompositions of those galaxies,
enabling us to include them at the low-mass end of the $M_{\rm BH} -
L_{\rm bulge}$ relationship for AGNs.  Recent reverberation-mapping
experiments that were carried out at MDM Observatory (spring 2007) and
Lick Observatory (spring 2008) focusing on AGNs with black hole masses
in the range $1 \times 10^6 - 5 \times 10^7 M_{\odot}$ (Denney et al.\
in preparation, Bentz et al.\ in preparation) show promise in further
extending the range and coverage of the $M_{\rm BH} - L_{\rm bulge}$
relationship for AGNs at the low-mass end.

\acknowledgements 

We would like to thank Alessandro Marconi and Tod Lauer for helpful
comments, and Chien Peng for his excellent program Galfit and for
helpful conversations regarding the galaxy fitting.  This work is
based on observations with the NASA/ESA {\it Hubble Space Telescope}.
We are grateful for support of this work through grants {\it HST}
GO-9851, GO-10516, and GO-10833 from the Space Telescope Science
Institute, which is operated by the Association of Universities for
Research in Astronomy, Inc., under NASA contract NAS5-26555, and by
the NSF through grant AST-0604066 to The Ohio State University.
M.B. gratefully acknowledges support from the NSF through grant
AST-0548198 to the University of California, Irvine, and
M.V. gratefully acknowledges support from {\it HST} GO-10417 and {\it
HST} AR-10691.  This research has made use of the NASA/IPAC
Extragalactic Database (NED) which is operated by the Jet Propulsion
Laboratory, California Institute of Technology, under contract with
the National Aeronautics and Space Administration and the SIMBAD
database, operated at CDS, Strasbourg, France.


\end{document}